\documentclass[prd,preprint,tightenlines,floatfix,showpacs,preprintnumbers,nofootinbib,eqsecnum]{revtex4}
 \usepackage[dvips,final]{graphicx}
  \usepackage{amssymb}
   \usepackage{amsmath}
    \usepackage{amsfonts}
     \usepackage{epsfig}
      \usepackage{bm}% bold math
      \usepackage{ulem}

\begin{document}

%\onecolumngrid
\begin{flushright}
LU TP 13-17\\
USM-TH-312\\
NSF-KITP-14-179\\
March 2015
\end{flushright}
%\twocolumngrid

\title{Quasi-classical Gravity effect on neutrino oscillations 
in a gravitational field of an heavy astrophysical object}

\author{Jonathan Miller}
\email{Jonathan.Miller@usm.cl}
 \affiliation{
Departamento de F\'{\i}sica Universidad T\'ecnica Federico Santa
Mar\'{\i}a\\ Casilla 110-V, Valpara\'iso, Chile}

\author{Roman Pasechnik}
\email{Roman.Pasechnik@thep.lu.se}
 \affiliation{
Theoretical High Energy Physics, Department of Astronomy and
Theoretical Physics, Lund University, S\"olvegatan 14A, SE 223-62
Lund, Sweden}


\begin{abstract} 
In the framework of quantum field theory, a graviton
interacts locally with a quantum state having definite mass, i.e.
the gravitational mass eigenstate, while a weak boson interacts with
a state having definite flavor, i.e. the flavor eigenstate. An interaction
of a neutrino with an energetic
graviton may trigger the collapse of the neutrino to a definite mass
eigenstate with probability expressed in
terms of PMNS mixing matrix elements. 
Thus, gravitons would induce quantum decoherence of a coherent neutrino flavor
state similarly to how weak bosons induce quantum decoherence of a
neutrino in a definite mass state. 
We demonstrate that such an
essentially quantum gravity effect may have strong consequences for
neutrino oscillation phenomena in astrophysics due to relatively
large scattering cross sections of relativistic neutrinos undergoing
large-angle radiation of energetic gravitons in gravitational field
of a classical massive source (i.e. the quasi-classical
case of gravitational Bethe-Heitler scattering). This
graviton-induced {\it decoherence} is compared to {\it decoherence} due to 
propagation in the presence of the Earth matter effect. Based on this
study, we propose a new technique for the indirect detection of
energetic gravitons by measuring the flavor composition of
astrophysical neutrinos.
\end{abstract}


\maketitle

% \vspace{-0.4cm}
\section{Introduction}
% \vspace{-0.2cm}
A theoretical extrapolation of the fundamental Quantum Mechanics
concepts to Einstein's gravity suffers from major difficulties with
quantization of space-time, ultraviolet behavior and
non-renormalizability of the resulting theory (for more details, see
Ref.~\cite{Christian,Woodard:2009ns} and references therein).
A wealth of theoretical studies have been presented in the
literature and many different quantum gravity models have been
developed. However, no conclusive statement about the true quantum nature of
gravity has been made. Only a real experiment can settle
the longstanding confusion between the different approaches and
provide guidance in developing the correct underlying theory.

Typically, in the standard quantum field theory framework which
unifies three of four basic forces of Nature, the
quantum gravity effects are disregarded as being
phenomenologically irrelevant at energy scales much smaller than the
Planck scale, $M_{Pl}\sim 10^{19}$ GeV. Moreover, due to enormous
suppression, quantum gravity effects are often referred to as nearly
unobservable \cite{Dyson,Rothman:2006fp}. While observing a single
graviton directly may be impossible, it is not impossible to find an
indirect evidence for quantum gravity. For an overview of potential
phenomenological opportunities for indirect signatures of quantum
gravity, see Refs.~\cite{AmelinoCamelia:2004hm,Damour:2008ji,
Hossenfelder:2010zj,Krauss:2013pha}. Nevertheless, our understanding of the quantum
nature of gravity suffers from the lack of accessible sources of
information.

In this paper, we propose a new approach for indirect experimental
studies of (local) quantum gravity interactions based upon an effect
of the large-angle energetic gravitational Bremsstrahlung (or
Gravi-strahlung, in short) off an astrophysical neutrino
passing through an external classical
gravitational potential on neutrino oscillation observables. This
process, known as the gravitational Bethe-Heitler (GBH)
process, can be considered in the quasi-classical approximation for
large angle and/or large energy graviton emission i.e. the Born
approximation is sufficient. Such a process may happen with a rather
high probability, such as in the case of an astrophysical neutrino scattering
off a massive source of classical gravitational field (like a star, black hole, dark 
matter distribution, or galaxy). In Quantum Mechanics, the latter process may serve
as a direct {\it quantum measurement} of the microscopic properties
of the gravitational field at astrophysical scales.

\subsection{Quasi-classical gravity}
In the limit of weak gravity, the quasi-classical approximation to
quantum gravity is a valid framework. In this case, the graviton
field is a correction determined on the flat Minkowskian background
and the metric operator in the Heisenberg representation is given by
$\hat{g}_{\mu\nu}=\eta_{\mu\nu}+\hat{h}_{\mu\nu}$. Here, the
$c$-number part $\eta_{\mu\nu}$ is the Minkowski metric and
$\hat{h}_{\mu\nu}$ is the graviton arising after the quantization
procedure. The Einstein-Hilbert action provides the mechanism for
{\it virtual gravitons} to propagate in the flat space-time and to
interact with one another in the quantum case as an analog of the
standard QED picture of the Coulomb field around an electric charge.
These virtual gravitons should be distinguished from {\it real
gravitons} which are radiated off an accelerated massive body and
their coherent wave packets correspond to gravitational waves in the
classical limit. A ``cloud'' of virtual gravitons around a static
massive body can be reinterpreted geometrically in terms of a
deviation from the flat metric (or curvature) in Einstein's
classical relativity \cite{Weinberg}\footnote{The background must be
chosen to be flat since only in this case is it possible to use the
Casimir operators of the Poincar\'e group and show that the quanta
have spin two and rest mass zero, thus being identified as
gravitons.}.

A graviton couples to the full energy-momentum tensor.
From the quantum-mechanical point of view, we work in the mass
eigenstate basis where the Hamiltonian of local
quantum-gravitational interactions has a diagonal form and identify
the particle mass eigenstates with gravitational eigenstates (due to
equivalence of gravitational and inertial mass). In this approach,
higher Fock states are created by the graviton creation
operator acting on a particle mass eigenstate. By measuring the
quasi-classical graviton cross section and deviations from it, we would be engaging in the 
first investigations of the deeper quantum gravity theory similar to how 
electroweak $\nu$-A measurements provided the first investigation of the deeper quantum
Weinberg-Salaam theory.

% \vspace{-0.4cm}
\section{Decoherence of neutrino state}
% \vspace{-0.2cm}
Generically, weakly-interacting neutrinos can be considered as an
efficient carrier of information across the Universe as they are not
absorbed or scattered by interstellar mediums. In practice, this
unique property of neutrinos enables us to utilize them for
large-scale astrophysical ``experiments'', such as searching for
possible tiny signatures of Lorentz invariance violation
\cite{Kostelecky:2011gq}, testing General Relativity
\cite{Fogli:1999fs} and Quantum Mechanics
\cite{Bahrami:2013hta,Ma:1998fp,Raghavan:2012sy}, testing the
equivalence principle \cite{Guzzo:2001vn,Anchordoqui:2005is}, testing
minimal length models \cite{Sprenger:2011jc,Sprenger:2010dg}, etc.
Ultimately, it is possible to identify an {\it
extraterrestrial large-scale quantum experiment} where neutrinos
 ``change'' their quantum state due to a
local quantum gravity process (in terms of local graviton coupling
to a fundamental matter particle) and further convey information
about such a process unchanged through the cosmological medium to
the Earth.
%\vspace{-0.4cm}
\subsection{Propagation decoherence}
% \vspace{-0.2cm}
The traditional source of decoherence typically referred to in
astrophysical neutrino oscillations studies can be called {\it
propagation decoherence}. This is when the distance that a neutrino
travels exceeds the neutrino oscillation length. In this case, the
neutrino mass states have separated so that they no longer interfere
at large distances from the production point. This source of
decoherence depends on the energy resolution of the detection
process, the energy of the neutrino, the masses of the neutrino mass states,
and other details of the production and detection processes.
In neutrino experiments, the time between neutrino production and
detection is normally not measured. In a real experiment, this means
that beyond the neutrino oscillation length the propagating neutrino
mass states no longer interfere during the interaction process in a
detector~\cite{Beuthe:2001rc,Giunti:1997wq}. For cosmic/astrophysical
neutrinos, in some cases and for some processes, this decoherence effect
is irrelevant~\cite{Kersten:2013fba}\cite{Farzan:2008eg}.

%%I could put in equations for Beuth (page 72ish)

% \vspace{-0.4cm}
\subsection{Classical Di\'osi-Penrose decoherence}
% \vspace{-0.2cm}
The role of classical Einstein's gravity in Quantum
Mechanics is under extensive consideration in the literature, and
may be sizeable under certain conditions. As was claimed in
Ref.~\cite{Christian:2005qa}, the gravity-induced quantum state
reduction can be tested by observing the neutrino flavor
oscillations at cosmological distances, while in
Ref.~\cite{Donadi:2012nt} it was regarded as practically
undetectable. This classical gravity effect on real-time evolution
of a quantum state composed of several mass eigenstates was
initially considered by Di\'osi \cite{Diosi} and Penrose
\cite{Penrose:1996cv}. In the classical gravity limit, the latter
can be approximated by a change in the phase of the flavor wave
function which appears mainly due to a non-degeneracy of neutrino
mass eigenstates, i.e. $\Delta m_{ij}^2\equiv m_j^2-m_i^2\not=0$,
where $m_{j}$ is the mass of the mass eigenstate $j$. This is caused
by different mass states traveling along different geodesics in
curved space-time and the whole effect gradually accumulates over
large cosmological distances~\cite{Ahluwalia:1998jx}. This is the
essence of {\it classical decoherence} of a neutrino flavor state
which is typically regarded as a probe for neutrino wave function
collapse models and, more generally, alternatives to conventional
(linear) Quantum Mechanics \cite{Bassi:2003gd}. Instead, we consider
another possible decoherence mechanism of a neutrino flavor state
triggered {\it at the quantum level} by a single local
graviton-neutrino interaction. Let us discuss this phenomenon in
detail.
% \vspace{-0.4cm}
\subsection{Quantum decoherence}
% \vspace{-0.2cm}

We expect elementary particles in the mass basis to be
gravitational eigenstates of the Hamiltonian of
quantum-gravitational interactions in the same way as leptons and
quarks are weak eigenstates in the flavor and CKM basis,
respectively. The advantage of the neutrino which we exploit here is
that they interact via the weak force, neutrino mass and flavor
eigenstates are not the same, and that they propagate at cosmological
distances/times. For particles whose flavor and mass eigenstates are
identical this technique would not work to identify that a graviton induced
quantum mechanical interaction had happened, which means that
the neutrino is a unique carrier of astrophysical quantum gravity interactions.

Consider first a relativistic neutrino state propagating in
the gravitational potential of a supermassive black hole, dark matter halo, or
other massive system. These are not only sources of strong
gravitational fields but could also be significant sources of
astrophysical neutrinos. Suppose now that at the quantum level a
graviton interacts only with a definite mass state (or gravitational
mass eigenstate) $a=1,2$ or $3$. 
This is equivalent to saying that
definite mass eigenstates (the propagating states)
are conserved by the quantum gravity
hamiltonian while superpositions, such as the flavor
eigenstates, are not \cite{Coriano:2013msa}. 
Note, the astrophysical neutrinos are
initially produced in electro-weak processes (e.g. in SNe processes) in a
definitive flavor state, $f=e,\mu$ or $\tau$, which are coherent
superpositions of mass eigenstates. In an astrophysical environment,
a high-energy graviton can interact only with a definite mass
component of the neutrino wave function thus causing {\it
quantum decoherence} of the neutrino which is in a
superposition of mass states, effectively ``converting''
it into a definitive mass eigenstate. This neutrino is quantum
mechanically observed as being in a definite mass state. This
means that between the production in an AGN or SuperNova (SNe)
or other astrophysics source and the detection in an Earth based
detector, the neutrino which was observed by the graviton 
exists in a definite mass state. This is independent 
from propagation decoherence.

The neutrino is ``converted" to mass state with a
probability $P_{\nu_f\to \nu_a}=|\Psi_{\nu_f\to \nu_a}|^2,$ given in
terms of the corresponding wave function $\Psi_{\nu_f\to \nu_a}$
which projects out a flavor state $\nu_f$ onto a mass state $\nu_a$
and is typically expressed in terms of the corresponding PMNS mixing
matrix element, $\Psi_{\nu_f\to \nu_a}\equiv V_{af}$. The
considering effect is different from other known classical
decoherence sources emerging due to a mere propagation (without a
hard graviton radiation) in classical gravitational potential and/or neutrino
propagation in flat space-time. 
The effect under consideration is a
straightforward consequence of fundamental time-energy uncertainty
relation for the real hard Gravi-strahlung and should be taken into account 
in studies of astrophysical neutrino oscillations.

The amplitudes of typical quasi-classical gravity scattering processes which
may lead to the quantum decoherence effect under certain conditions
can be represented as follows:
\begin{eqnarray*}
&&A^{(G),1}_{\nu_f\to \nu_a}=\Psi_{\nu_f\to \nu_a}A^{(G)} (\nu_a + G
\to
\nu_a + G)\,. \\
&&A^{(G),2}_{\nu_f\to \nu_a}=\Psi_{\nu_f\to \nu_a}A^{(G)} (\nu_a + M
\to \nu_a + G + M)\,.
\end{eqnarray*}
Here, $M$ is a source of strong classical gravitational fields, such
as a massive star or a black hole. The first amplitude corresponds
to the gravitational Compton scattering of a neutrino mass state off
a real graviton in the medium, the second amplitude represents the
GBH scattering of a neutrino in gravitational mass state off a
classical heavy source $M$ (with energetic graviton radiation).
Clearly, a mass eigenstate $\nu_a$ ``produced'' in this interaction
due to decoherence does not undergo oscillation until it interacts
weakly with normal matter (e.g. in an Earth detector) by means of
$W,Z$-exchange. Therefore, quantum decoherence may have an
non-negligible effect on neutrino oscillation observables, along
with other existing sources of classical decoherence and medium matter
effects \cite{Mikheev:1986wj,Wolfenstein:1977ue}. 
%% below could go later
Explicitly, oscillation characteristics of neutrinos coming from e.g. a vicinity
of the Galactic Center may differ from vacuum oscillations. The
latter case could be where a source of neutrinos is ``nearby'' but
where there is no massive objects between the source and the Earth
(nor significant variations in dark matter density). Such neutrinos,
if identified, could be used as a control sample.

In a sense, the quantum gravity-induced decoherence of a definite
flavor state described above is in close analogy to the weak-induced
decoherence of a definite mass state.
For example, $W,Z$ bosons interact only with a coherent flavor state
inducing a ``conversion'' of a definite mass state into a definite
flavor state. Namely, a neutrino in a mass eigenstate $\nu_a$ turns
into a flavor eigenstate $\nu_f$ through an interaction with the
 virtual $Z,W$-bosons propagating in the $t$-channel, i.e.
four different reactions are possible
\begin{eqnarray*}
&&A^{\rm (w),1}_{\nu_a\to \nu_f}=\Psi_{\nu_a\to \nu_f} A^{\rm (w)}
(\nu_f
+ l_f' \to \nu_f' + l_f)\,, \\
&&A^{\rm (w),2}_{\nu_a\to \nu_f}=\Psi_{\nu_a\to \nu_f} A^{\rm (w)}
(\nu_f
+ l_f' \to \nu_f + l_f')\,, \\
&&A^{\rm (w),3}_{\nu_a\to \nu_f}=\Psi_{\nu_a\to \nu_f} A^{\rm (w)}
(\nu_f
+ N \to \nu_f + X)\,, \\
&&A^{\rm (w),4}_{\nu_a\to \nu_f}=\Psi_{\nu_a\to \nu_f} A^{\rm (w)}
(\nu_f + N \to l_f + X)\,,
\end{eqnarray*}
such that $\Psi_{\nu_a\to \nu_f}=\Psi_{\nu_f\to \nu_a}^*$. Here, a
definitive mass state which may exist due to previous hard
neutrino-graviton interaction or due to the resonance MSW effect
\cite{Mikheev:1986wj,Wolfenstein:1977ue} is ``converted'' back into
a flavor state which may undergo oscillation.
It is important to note that because the neutrino is not likely to
interact weakly between the source and the Earth, if the neutrino is in
a definitive mass state induced by the hard
neutrino-graviton scattering event which occurred long before it arrives at the
Earth it will still be in the definitive mass state at the earth. 
The distance between a hard neutrino-graviton scattering
event and detection event is not important.

In the case of vacuum neutrino oscillations, the traveling neutrino
is not in a definitive mass eigenstate but is rather in a
superposition of mass eigenstates which evolves when the neutrino
travels in space-time. Then, with respect to the weak interactions,
the non-diagonal $\Psi_{\nu_f\to \nu_{f'}}$ transition amplitude
between two flavor states $f$ and $f'$ is given by
\cite{Beringer:2012pdg}
\begin{eqnarray}
    \Psi_{\nu_f\to \nu_{f'}}=\sum_{j} V_{f'j} e^{-i\frac{m_j^2}{2E_\nu}L}
    V_{fj}^{*}\,,
\end{eqnarray}
here $L$ is the distance from where the neutrino was created in a
definite flavor eigenstate $\nu_f$, and $E_\nu$ is the energy of the
neutrino. Analogically, for neutrino-graviton interactions the
$\Psi_{\nu_f\to \nu_{a}}$ transition amplitude between a flavor
state $f$ and a mass state $a$ can be written as
\begin{eqnarray}
    \Psi_{\nu_f\to \nu_a}= e^{-i\frac{m_a^2}{2E_\nu}L}\,V_{af}\,
\end{eqnarray}
which means that the probability for a given flavor neutrino state
$f$ to decohere by transforming into a mass state $a$ due to a hard
graviton-neutrino interaction, given by $P^{(G)}_{\nu_f\to
\nu_a}\sim |A^{(G)}_{\nu_f\to \nu_a}|^2=|\Psi_{\nu_f\to
\nu_a}|^2|A^{(G)}|^2$, is independent of the neutrino mass, $m_a$,
the mass splitting, $\Delta m_{ab}$, and the distance from the
neutrino source, $L$. The dependence on the relativistic neutrino
energy, $E_\nu\gg m_a$, for a given scattering comes from the
neutrino mass state scattering amplitude squared, $|A^{(G)}|^2$ (for
more details, see the next Section).
%-------------------------------------------------------------
\begin{figure}[h!]
\centerline{\epsfig{file=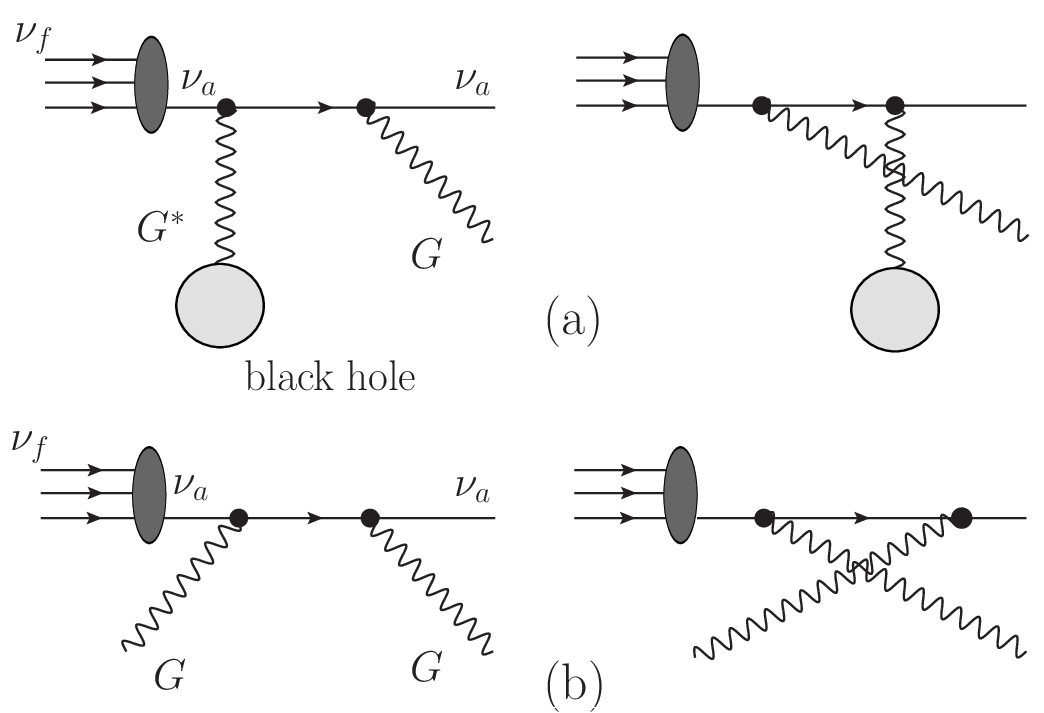,width=8cm}} \caption{The
quasi-classical gravity processes which destroys the coherence of the
neutrino flavor eigenstate ($f=e,\mu,\tau$) at the quantum level
effectively turning it to a mass eigenstate ($a=1,2,3$) -- the
gravitational Bethe-Heitler-type scattering of neutrino off a
massive object, e.g. a black hole (a), and the gravitational Compton
scattering (b). The dark ellipse is a projection to a fixed mass
state and the shaded circle is a classical source of the
gravitational field.}
\label{fig:BH-nu}
\end{figure}
%-------------------------------------------------------------

\subsubsection{Differences from other sources of decoherence}

Contrary to the Penrose-Di\'osi effect of classical decoherence
\cite{Diosi,Penrose:1996cv}, the quantum decoherence of a neutrino
flavor state happens at small space-time scales, $\Delta l_{dec}$,
which are much smaller than the neutrino wave length scale: $\Delta
l_{dec} \ll L_\nu$, due to the quantum nature of the tree-level 
graviton-neutrino interaction. An additional
significant difference, the quantum decoherence effect is not
sensitive to the mass differences of the mass eigenstates, or to
$\Delta m_{ij}^2$, while they are crucial for and determine
classical decoherence of the neutrino flavor state at large
separations, $\Delta l_{dec} \gg L_\nu$. Most importantly, quantum
decoherence provides us with a key for phenomenological verification
of quantum gravity models with possible deviations from
quasi-classical gravity through measurement of neutrino
oscillation characteristics.

\begin{figure}[h!]
\centerline{\epsfig{file=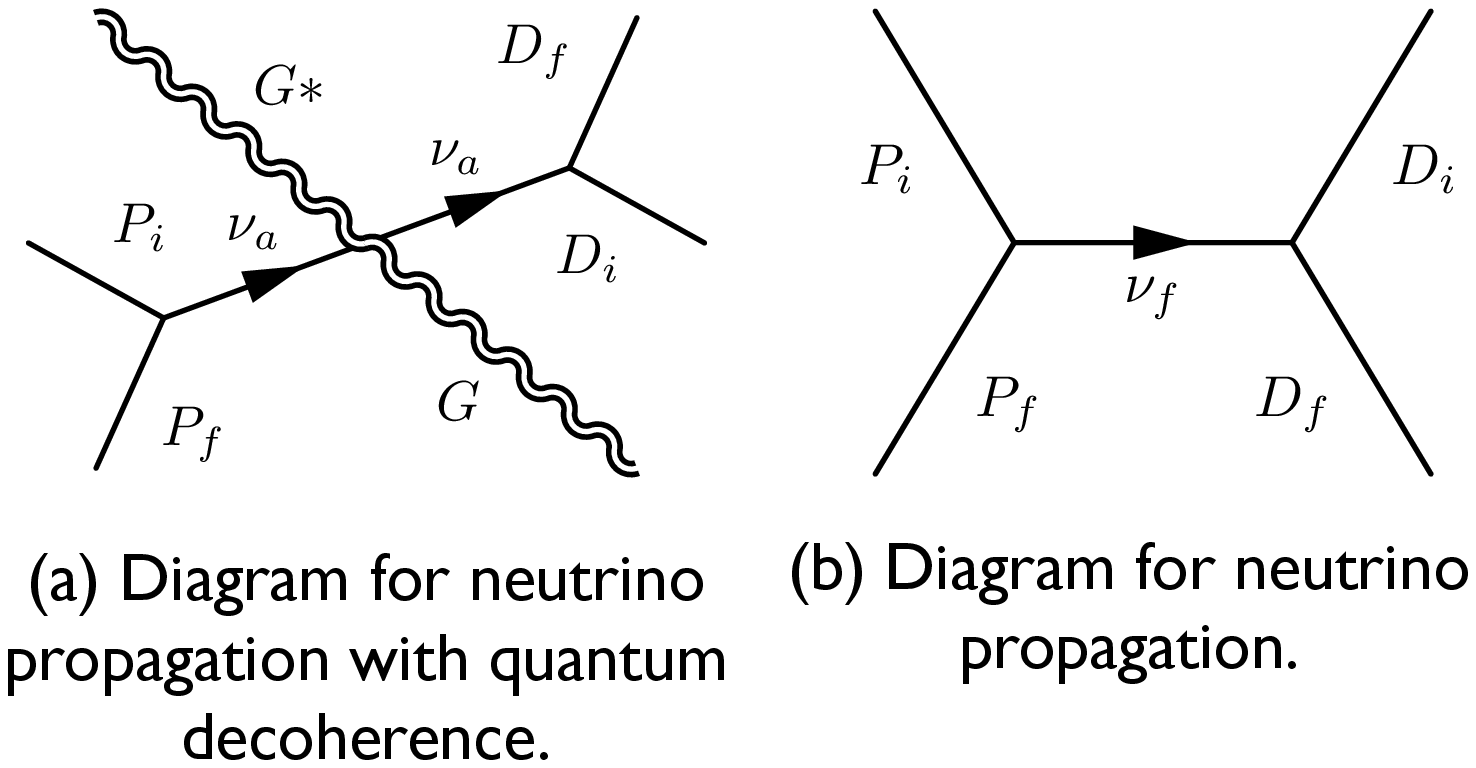,width=8cm,angle=270}}
\caption{Shown are diagrams of neutrino propagation in the quantum field
theoretical description (such as found in reference \cite{Beuthe:2001rc}). The neutrino is described
as a stretched propagator between the production ($P$) and detection ($D$) (subfigure b)
weak processes (with initial and final states denoted by subscripts $i$ and $f$) . In 
the case of quantum decoherence (subfigure a), the emission of a graviton in between the 
production and detection processes means that the neutrino exists in a definitive 
mass state since both the propagation hamiltonian and graviton interaction
hamiltonian conserve mass. The quantum decoherence effect described here follows 
from the discussion in section 2 of \cite{Beuthe:2001rc}.}
\label{fig:fig2diagram}
\end{figure}

The proposed effect is also different from the standard propagation
decoherence (see Fig. \ref{fig:fig2diagram}). In propagation decoherence, the neutrino mass states
are separated in time and/or space and so the local weak interaction
(the detection process) {\it observes} an incoherent sum of the
propagating mass states in a given space-time point. In quantum
decoherence, the neutrino exists only within a given mass eigenstate
after being ``observed'' by the hard graviton (e.g. in the quantum
processes of hard GBH or Compton scattering, see below). This
difference is important. Indeed, while a flux of neutrinos which
have undergone quantum decoherence is {\it observed} by a weak
interaction in an Earth-based detector as an incoherent sum of the
mass states, they do not experience a change of potential induced by
matter (for example, the MSW effect) as an incoherent sum of mass states. Namely,
the neutrino which has not undergone quantum decoherence
experiences matter as a superposition of mass states, while the neutrino which has
undergone quantum decoherence would not experience matter
as a superposition of mass states.
Also, it is possible that a neutrino passing through densities which change
non-adiabatically might demonstrate interference phenomena as
presented in Ref.~\cite{Kersten:2013fba}. As we will explicitly
demonstrate below, such a difference between the quantum and
propagational decoherences in the presence of the Earth matter effect
may be observable and is important in studies of astrophysical neutrino
oscillations.

\subsection{Decoherence in the presence of the Earth matter effect}
\label{sec:decohmatter}

We would like to note that while the flux from quantum (gravitational) decoherence is a flux of pure
mass eigenstates as noted, that the important difference is that in the propagation decoherence
case the flux is not of pure mass eigenstates, but rather decoherent (spatially separated) mass
eigenstates. No quantum measurement of the state of these neutrinos has taken place,
and the neutrino still exists as a superposition of mass states (just no longer with off diagonal elements
in the density matrix). While these
two situations are exactly the same when detected in the case
where the flux is detected without passing through matter; in the case where the flux passes
through matter, the regeneration which the neutrino flux experiences is different for the two cases.
In the quantum gravitational decoherence case, the neutrino flux experiences regeneration as
fluxes of neutrinos in pure mass eigenstates. However, in the propagation decoherence case, the
neutrino flux experiences regeneration as a superposition of mass eigenstates; individual actual
neutrinos continue to exist in a spatially separated quantum superposition of mass eigenstates.
These spatially separated quantum superpositions experience the potential of the Earth. 
Simulation was done to demonstrate the possible size of this effect due to the difference in 
regeneration in the two cases (details below). In the simulation the neutrino is considered to have experienced
propagation decoherence and the exact distance the neutrino travelled is not important (as long as 
it fulfils the conditions in \cite{Farzan:2008eg}); in this work we 
consider relatively small distances ($<$1000 kpc) since we do not explicitly 
consider differentiating quantum decoherence from potentially very 
long distance effects, for example classical decoherence.

The simulation presented here operates in the $S$-matrix oscillations formalism
\cite{Ohlsson:1999xb} and was realised in Python, 
but for the neutrino propagation the code gives the same results as
the Fortran simulation found in \cite{wimp}. In the case of quantum decoherence where the
neutrino exists as a single mass eigenstate, the neutrino is 
not coherent and the earth matter effect causes no significant change to
the measured flavor composition at low energies.
For the case of propagational decoherence, the neutrino is a decoherent
superposition of mass states where the description of the neutrino 
in the flavor basis is given by constant phase differences between the mass eigenstates
and an effect may be seen. While
many models for supernova neutrino spectra, see e.g. Refs. \cite{Gava:2009}\cite{Vaananen:2011}
and other astrophysical neutrino sources may be interesting, for
simplicity, we adopt an initial uniform electron neutrino flux
with energies between 0.5 and 20 MeV.  We are interested in a
significant measurement so incorporated propagation of the neutrino
directly through the Earth, and have obtained a maximal difference between propagation
decoherence and the quantum
decoherence effect at greater than 100\%. The latter is 
important for next generation neutrino measurements.\\

\subsection{Analytic illustration: two flavour case without core}

The theory of neutrino propagation, including neutrino propagation in medium and neutrino
propagation when the neutrino experiences propagation decoherence are well presented in
the papers by Beuthe \cite{Beuthe:2001rc} and Akhmedov and collaborators 
\cite{Akhmedov:2012mk,Akhmedov:2010ms} and Blennow and Smirnov \cite{Blennow:2013rca}. 
These papers give the essential understanding of neutrino propagation in matter and propagation 
decoherence, but no explicit formula is given for a neutrino which undergoes propagation 
decoherence and then experiences the Earth matter effect.

For ease of discussion we will consider just two regimes, the vacuum and the earth (with
constant density) and two neutrino flavors. Due to the discontinuity at the earth's surface,
the adiabatic formulas do not describe the neutrino propagation. However, the solution is
to match the flavor conditions between the two regimes. The flavor at the point
before the density jump is used to determine the initial state after the jump 
\cite{Blennow:2013rca,Balantekin:1996ag}.

Propagation decoherence was studied in detail by Beuthe \cite{Beuthe:2001rc} but unfortunately
only in the vacuum case. Akhmedov and Wilhelm \cite{Akhmedov:2012mk} explicitly consider decoherence
due to production or detection conditions but only notes that finite coherence length is recovered
during the integration over energy. Beuthe \cite{Beuthe:2001rc} goes into great detail about the
physics, which is that the wave packets separate or that the wave packet spatial spread is so large
that the phase varies over the wave packet and the information is lost. These are changes of the 
relationship between the states which make up the neutrino, and the changes can be accounted for 
by integrating the phase in the vacuum transition probability. In the case of two regimes with a sharp 
transition, as we consider here, it is necessary to find the flavor states at the transition.

The condition for the wave packet separation to be complete is given explicitly by Smirnov and 
Farzan~\cite{Farzan:2008eg}. They explicitly note that this is
different than the effect due to averaging (Section 2.1 in Ref.~\cite{Farzan:2008eg}) over the energy, despite the effect being
computationally the same for the vacuum (and adiabatic) case \cite{Beuthe:2001rc}. As noted by
Ref.~\cite{Farzan:2008eg}, once the phase difference becomes large, the phase difference between 
the mass eigenstates can be expressed as a constant. This happens once \cite{Farzan:2008eg} 
\begin{equation}
\sigma_{x} \ll d_{L} = 3\times 10^{-3} \mathrm{cm}\frac{L}{100 \mathrm{Mpc}} 
\frac{\Delta m^2}{2.5\times 10^{-3}\mathrm{eV}^2}\left(\frac{10 \mathrm{TeV}}{E}\right)^2 \,,
\end{equation}
which is achieved for both $\Delta m_{12}^{2}<1$ eV and $\Delta m_{23}^{2}<1$ eV for $L=10$ kpc since
we expect $\sigma_{x}$ to be less than $10^{-10}$.

To determine the proper state we must consider the proper normalisation and phase for the states.
In the two flavor approximation, the probability is given by
\begin{equation}
P_{ee}=\frac{1}{2}\left(1+\cos^{2}\left(2\theta\right)\right)
\end{equation}
and
\begin{equation}
P_{e\mu}=\frac{1}{2}\sin^{2}\left(2\theta\right)
\end{equation}
where $\theta$ is the neutrino two-flavor mixing angle in vacuum and $e$ and $\mu$ are the
two neutrino flavors. The amplitude can then be given by
\begin{equation}
A^{dec}_{ee}=\cos^{2}\left(\theta\right)e^{i \frac{3\pi}{4}}+\sin^{2}\left(\theta\right)e^{-i \frac{3\pi}{4}}
\end{equation}
and
\begin{equation}
A^{dec}_{e\mu}=\sin\left(\theta\right)\cos\left(\theta\right)\left(e^{-i\frac{3\pi}{4}}-e^{i \frac{3\pi}{4}}\right)~.
\end{equation}
These give the correct flavor amplitudes of a neutrino produced in a $\nu_{e}$ state which
has travelled through vacuum and experienced wave packet separation when it reaches the
earth-to-vacuum transition. There is an overall phase, but for length scales (such as the earth-to-vacuum
transition) much smaller than the the
distance between the wave packets (which can be 1 km or more) the phase difference between the 
wave packets is a constant as expressed above. An amplitude where the phase between the wave packets changes with
distance would be incorrect for large wave packet separations.

This allows us to give a clear description of a produced $\nu_{e}$ which experiences propagation
decoherence, travels through the mantle of the earth, and then is detected as a $\nu_{e}$. This is
\begin{equation}
P^{prop}_{ee}=|A^{dec}_{ee}A^{mat}_{ee}+A^{dec}_{e\mu}A^{mat}_{\mu e}|^{2}
\end{equation}
which for the standard description in terms of a matter mixing angle $\theta_{m}$ and the matter
phase $x_{m}$ is
\begin{eqnarray} \nonumber
P^{prop}_{ee}&=&\frac{1}{8}\Big(2\cos^{2}\left(x_{m}\right)\left(3+\cos\left(4\theta\right)\right)
+\sin^{2}\left(x_{m}\right)\Big(4+\cos\left(4\theta_{m}\right)+\cos\left(4\theta_{m}-8\theta\right) \\
&+& 2\cos\left(4\theta_{m}-4\theta\right)+\sqrt{8}\sin^{2}\left(\theta\right)
\sqrt{3+\cos\left(4\theta\right)}\sin\left(4\theta_{m}-4\theta\right)\Big)\Big)\,,
\end{eqnarray}
where\cite{Ohlsson:1999xb}
\begin{eqnarray}
&&\sin\left(2\theta_m\right)^2=\frac{\sin\left(2\theta\right)^2}{\sin\left(2\theta\right)^2+
\left(\cos\left(2\theta\right)-\frac{2 A E_{\nu}}{\Delta m^2}\right)^2} \,, \\
&& x_m=x\sqrt{\sin\left(2\theta\right)^2+\left(\cos\left(2\theta\right)-\frac{2 A E_{\nu}}{\Delta m^2}\right)^2}\,,
\qquad  x = \frac{\Delta m^2 L}{4 E_{\nu}} \,,
\end{eqnarray}
and $A$ is a constant density.
This formula is different than that which is given for solar neutrinos and which is presented in the 
paper by Dighe, Liu, and Smirnov \cite{Dighe:1999id}. They give the calculation for an incoherent mixture
of mass eigenstates originating in the sun, but this important paper does not explicitly consider propagation 
decoherence (decoherence due to wave packet separation) but rather the effects of coherent neutrino
propagation at long baselines in addition to the known solar MSW resonance effect where neutrinos
which leave the sun exist in only a single mass eigenstate. Smirnov and Farzan \cite{Farzan:2008eg} explicitly
give the condition to have the phase change for decoherent, and also for detection to restore coherence.
The condition is that the measurement takes place over large enough time scales (or small enough energy 
resolutions) for both wave packets to be measurable \cite{Kiers:1995zj}. 

In the case of quantum decoherence we have the emission of a graviton off a neutrino mass state in the vacuum. 
An interaction of a neutrino with the graviton serves essentially as a measurement of the neutrino state, 
both its detection and production in quantum-mechanical language.
This tells us the condition on the graviton which must be true for the effect presented in this work, it is the 
condition of coherent production/detection of a neutrino as presented in Refs.~\cite{Beuthe:2001rc} and 
 \cite{Akhmedov:2012mk}. The formulation can be considered in the framework of Ref.~\cite{Akhmedov:2012mk} but where the $\tilde{U}$ matrix elements
 for the mass eigenstates and not for the matter eigenstates (so the identity since we are assuming propagation in the
 vacuum). Then the relationship for neutrinos which have undergone quantum gravity decoherence is simply (for two flavours):
 \begin{equation}
 P_{\alpha\beta} =  P_{\alpha 1}P^{earth}_{1\beta} + P_{\alpha 2}P^{earth}_{2\beta}~.
 \end{equation}
 This is equivalent, as discussed in \cite{Dighe:1999id}, to a coherent neutrino being observed a long
 distance from the neutrino source by an experiment without arbitrary energy resolution. The probability is then
 given by:
\begin{eqnarray} \nonumber
P^{grav}_{ee}&=&\frac{1}{16}\Big(10+2\cos\left(4 \theta_{m}\right) - \cos\left(4\theta_m-2 x_m\right)+2\cos\left(2 x_m\right) \\
\nonumber &-& \cos\left(4\theta_m+2 x_m\right)+4\cos\left(4\theta\right)\left(\cos^2\left(2\theta_m\right)+
\cos\left(2 x_m\right)\sin^2\left(2\theta_m\right)\right) \\
&+&4\sin\left(4\theta_m\right)\sin^2\left(x_m\right)\sin\left(4\theta\right)\Big)\,,
\end{eqnarray}
where $x_m$ and $\theta_m$ are as before.

The ratio of neutrinos which have undergone {\it propagation decoherence} and at the same time propagated 
through a region of constant density to those which have only propagated through the vacuum is given
by the following expression
\begin{eqnarray} \nonumber
R_{p}&=&\Big(\cos\left(x_m\right)^2\left(3+\cos\left(4\theta\right)\right)+
\left(2+\cos\left(4\theta_m-8\theta\right)+\cos\left(4\theta_m-4\theta\right)\right)\sin\left(x_m\right)^2 \\
&-&\, 2\sin\left(2x_m\right)\sin\left(2\theta_m-2\theta\right)\sin\left(2\theta\right)\Big)/(3+\cos\left(4\theta\right)) \,.
\end{eqnarray}

Analogically, the ratio of neutrinos which have undergone {\it quantum decoherence} in the presence of matter 
effect in a constant density medium to those which have propagated through the vacuum takes a different form
\begin{eqnarray}
R_{q}=\frac{5+\cos\left(4\theta_m\right)+\cos\left(4\theta_m-4\theta\right)+
\cos\left(4\theta\right)+4\cos\left(2x_m\right)\cos\left(2\theta\right)
\sin\left(2\theta_m-2\theta\right)}{6+2\cos\left(4\theta\right)} \,.
\end{eqnarray}
A difference between the ratios $R_p$ and $R_q$, in principle, could be measurable and 
indicates the principal difference between propagation and quantum decoherences emerging
in the presence of matter effect. Measurement of such a difference could therefore 
serve as a clear example of graviton detection. 

%%In the presence of an additional jump in matter density (or a core) the analytical analysis becomes too involved to be shown 
%%here, but the corresponding numerical results are presented in \cite{}.\\

\section{Graviton-neutrino scattering}

Now consider which quantum gravity processes the neutrino could
possibly experience so as to experience the quantum decoherence
effect in the astrophysical medium. As mentioned we will be considering 
quasi-classical gravity processes.

As is known the Coulomb field is measured by inserting a charged
probe into it. From the quantum electrodynamics (QED) point of view,
an electromagnetic scattering of a charged particle off the Coulomb
field is due to an exchange of virtual photons (with small negative
momentum transfer squared $-q^2=Q^2>0$ in the $t$-channel) between
the probe and the source. Analogically, it is correct to discuss
multiple exchange of virtual $t$-channel gravitons in a scattering
event as a signature of non-zeroth curvature itself (for more
detailed discussions of the principles, see e.g.
Ref.~\cite{Boulware}).

Generically, in quantum electrodynamics (QED) the virtual photons
may become real (produced on-mass-shell) if one disturbs the field
pumping energy into it. This is the physical reason for photon
Bremsshtrahlung in QED. Specifically, the standard Bethe-Heitler
scattering in electrodynamics demonstrates that only an accelerated
charge emits real photons (corresponding to electromagnetic wave in
the classical limit of multiple soft photon radiation). Likewise, in
the quasi-classical gravity framework, the virtual graviton, as a
quantum of the gravitational field of a static massive object, may
turn into the real one (corresponding to gravitational wave in the
classical limit of multiple soft graviton radiation) if the source
of the gravitational field is accelerated or, in general, when the
energy-momentum tensor experiences disturbances.

Possible sources of real gravitons in the Universe include: active
galactic nuclei (AGN), binary systems, SuperNova explosions (SNe),
primordial black holes collisions, compact star/black holes
binaries, quantum bremsstrahlung of gravitons of particles
scattering off a massive object, black hole (BH) evaporation, relic
isotropic gravitational background from the early universe,
inflation, phase transitions in the primordial plasma, the decay or
interaction of topological defects (e.g. cosmic strings), etc. For
details and references, see Ref.~\cite{Dolgov}.

Consequently, in the cosmological medium a neutrino can scatter
either off a classical gravitational potential with accompanying
radiation of an energetic real graviton off the scattered neutrino
(e.g. Bethe-Heitler-type scattering) or off real graviton in the
astrophysical medium (e.g. Compton-type scattering). Let us consider
both cases and conditions for initiation of the quantum neutrino
decoherence in more detail.

% \vspace{-0.4cm}
\subsection{Gravitational Bethe-Heitler scattering}
% \vspace{-0.2cm}

In fact, all elementary particles, including neutrinos, when
traveling in the vicinity of massive objects (sources of classical
gravitational field) can emit real gravitons with a certain energy
spectrum. This process has a straightforward QED analog of a photon
 emission in relativistic electron scattering off the Coulomb field of
a heavy nucleus mentioned above, the Bethe-Heitler process at the
Born level. Even though the energy spectrum of radiated real
gravitons is peaked in the forward direction and in the
infrared limit (corresponding to forward radiation of classical
gravitational waves), there is a non-negligible probability to
radiate {\it hard or energetic gravitons}, namely, with energies
comparable to the incoming relativistic neutrino energy. Due to the
quantum nature of the neutrino and graviton, the latter process can
trigger a dramatic decoherence of an incoming neutrino flavor state
at the quantum level during a very short time scale (inversely
proportional to the energy of the radiated hard graviton).

The decoherence of the neutrino at the quantum level can only be
initiated by hard energetic interactions with relatively hard
gravitons whose energies exceed the mass difference between
different mass states $E_G\gtrsim \Delta m_{ij}$ and therefore
requires hard real graviton emission. In this case, the hard
graviton probe has a small wave length and thus can resolve separate
mass states in a coherent or incoherent neutrino flavor state in the
quantum mechanical sense\footnote{Likewise a hard enough photon can
resolve an internal substructure of the proton wave function and
interacts with separate quarks it is composed of while a soft one
``sees'' a proton as a whole only.}.

A soft graviton with energy lower than the difference between
mass states will be unable to resolve the individual mass
eigenstates in this superposition and will instead couple to the
whole energy-momentum tensor of the flavor state, non-locally, which
is the classical General Relativity limit. In the latter case
quantum decoherence is not triggered, and the effect will be as discussed
in \cite{Ahluwalia:1998jx}.

The Born-level calculation is good first order approximation
in the case of off-forward
hard graviton emissions at large angles relevant for the quantum
decoherence effect -- this is the reason why one can disregard
higher-order radiative corrections which are highly suppressed 
(by extra powers of the Planck mass) as long as one cuts
off the problematic but uninteresting infrared/collinear parts of
phase space. As was previously shown in
Ref.~\cite{Barker}, the radiative corrections can only be relevant
in the deep infrared limit of soft real gravitons $E_G\to 0$ emitted
in the forward direction where they will cancel the soft/collinear
divergences. The latter classical limit represents classical
gravitational waves emitted off a neutrino state with very
small or no impact on it.

In the considering GBH case, shown in Figure~\ref{fig:BH-nu}(a), one
deals with the graviton exchange with negative momentum transfer
squared $t=-q^2<0$ in the $t$-channel with the propagator stretched
between the relativistic neutrino of mass $m_\nu$ and energy
$E_\nu\gg m_\nu$ and a massive classical gravitational field source
with mass $M\gg E_\nu$. The wave function, $\Psi_{\nu_f\to \nu_a}$,
describes a projection of a given flavor state $f$ onto a fixed mass
state $a$ is denoted as a dark ellipse, while the heavy classical
source of the gravitational field is shown by a shaded circle.
%-------------------------------------------------------------
%\onecolumngrid
\begin{figure*} %%[h!]
\begin{minipage}{0.325\textwidth}
 \centerline{\includegraphics[width=1.1\textwidth]{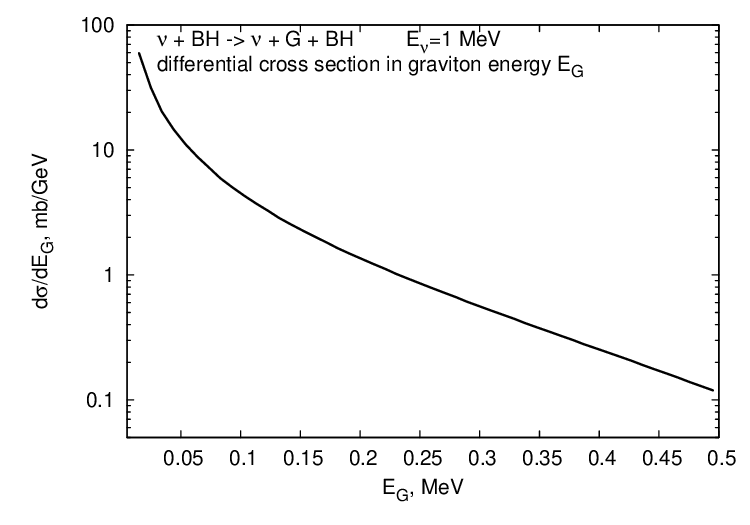}}
\end{minipage}
\begin{minipage}{0.325\textwidth}
 \centerline{\includegraphics[width=1.06\textwidth]{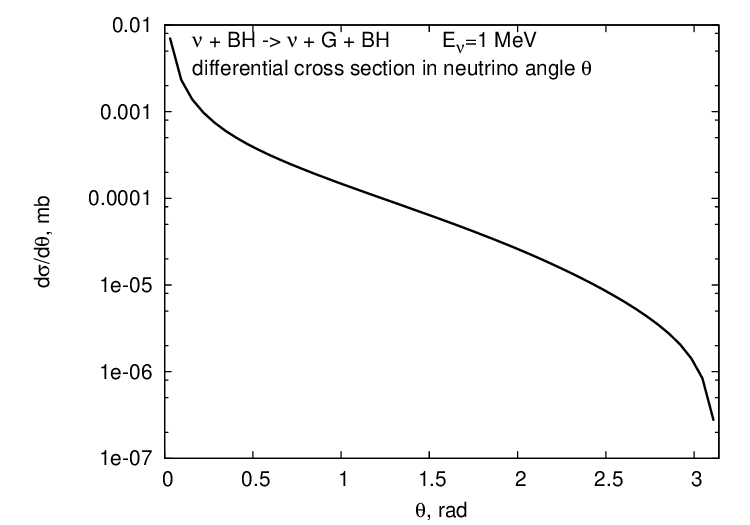}}
\end{minipage}
\begin{minipage}{0.325\textwidth}
 \centerline{\includegraphics[width=1.1\textwidth]{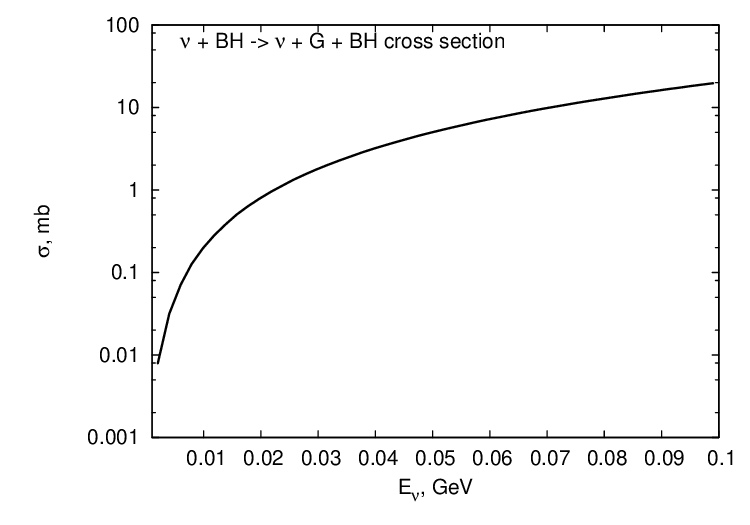}}
\end{minipage}
\caption{Differential cross section of the gravitational
Bethe-Heitler scattering of neutrino off a massive object e.g. a
black hole (BH) in radiated graviton energy $E_G$ (left), in polar
angle of the final-state neutrino $\theta_{\nu}$, and the integrated
cross section as a function of incoming neutrino energy $E_\nu$
typical for astrophysical sources, e.g. see Ref.~\cite{Janka:2012wk}
(right).} \label{fig:CS-GBH}
\end{figure*}
%\twocolumngrid
%-------------------------------------------------------------

The GBH cross section has initially been calculated for the
gravitational scattering of scalar particles with $M\gg m$ in
Ref.~\cite{Barker}. In the soft graviton limit, the
graviton-neutrino coupling is not sensitive to the spin of an
incident relativistic particle to leading order, while the classical
non-relativistic source can be considered to be spinless in this
first discussion for simplicity (in principle, helicity dependence
of hard graviton-neutrino interactions can be a relevant topic for
further studies). We therefore use their formula as a sufficiently
good approximation to estimate the neutrino-solar mass cross section
with graviton radiation numerically. In this case, as an order-of-magnitude estimate, the GBH
cross section at the Born level behaves as
\begin{eqnarray} \label{GBH-CS}
\sigma_{\rm GBH}\sim \frac{M^2E_{\nu}^2}{M_{Pl}^6}\,, \quad M\gg
E_{\nu}\gg m_{\nu}\,,
\end{eqnarray}
and thus may not always be very small since the Planck scale
suppression can be largely eliminated by having a mass $M$ of a
heavy classical source in numerator. In particular, for a solar mass
object $M\sim 10^{57}$ GeV, we have $M^2/M_{Pl}^6\sim 1$ GeV$^{-4}$,
so there is no significant suppression of the cross section for
relativistic neutrinos. This is a particle physics magnitude cross section
which naively implies particle physics size impact parameters.
Larger impact parameters would exist for
larger masses, such as the dark matter halo. The above cross section 
is integrated over impact parameter, and it may be instructive to look 
into differential cross section in order to find the probability of this 
process as a function of distance to the massive astrophysical body. 
For the current study it suffices to note that such probability can 
potentially be significant.

It is worth noticing that, the Bethe-Heitler calculation in QED to
first order gives the correct cross section for the photon
Bremsshtrahlung for extended objects such as a nucleus as shown in
Ref.~\cite{Tseng} (Ref.~\cite{Amundsen} demonstrates that after
advances it is still correct to first order in the hard photon
limit). Similarly, one may expect that the GBH result for a
point-like classical source of gravitational potential should be
roughly correct to first order for extended objects as well, like a
star or even a dark matter distribution in the quasi-classical
gravity case. The QED calculations are provided in above references
for scattering due to the field surrounding the nucleus, the nucleus
is small compared to the total size of the field and scattering off
the nucleus via other processes can be safely ignored. Most of the
cross section is thus not due to trajectories where the electron
passes through the nucleus. We expect that this is also true for the
GBH scattering, where most of the flux which scatters off the star
or other massive object will not pass through the star, and the
point-like estimation (\ref{GBH-CS}) remains valid.

In Figure~\ref{fig:CS-GBH} we have presented the differential (in
radiated graviton energy $E_G$ and neutrino angle $\theta_\nu$) and
integrated cross sections of the GBH process for typical MeV-scale
astrophysical neutrinos and a solar mass scale source of the
gravitational field. As expected, the main bulk of the cross section
comes from the soft gravitons (gravitational waves) emission in the
forward limit. It is remarkably important, however, that there is a
long non-negligible tail in the differential distributions of the
GBH cross section in the single real graviton energy $E_G$ and
emission angle $\theta$. It turns out that such a tail to
harder/off-forward gravitons is not very strongly suppressed --
typical GBH scattering cross sections for SNe neutrino energies of
$E_{\nu}\sim 10-100$ MeV and a Solar-mass classical source are found
to be around $\sigma\sim0.1-10$ millibarns, which are some $16-18$
orders of magnitude larger than typical neutrino-electron scattering
cross sections (less than an attobarn at the same energies).

This observation strongly suggesting the importance of the quantum
decoherence initiated by interactions with such energetic real
gravitons. The latter source of decoherence does not have a
classical interpretation. As we have already mentioned, due to a
quantum-mechanical nature of a {\it single hard graviton} emission
at energies $E_G\gtrsim \Delta m_{ij}$ (with local coupling to a
gravitational mass eigenstate) and universal quantum mechanical
time-energy uncertainty arguments, the considered effect of neutrino
flavor decoherence is a purely {\it quantum effect}. The hard
Gravi-strahlung effect is thus relevant for a broad range of
neutrino energies, and one can utilize the SNe neutrinos as a clear
sample since (1) fluxes of SNe neutrinos are the largest among
astrophysical neutrinos and (2) SNe neutrino emission mechanisms are
the best understood among other possible astrophysical sources.

Of course, the $t$-channel gravitons are extremely soft and form
classical gravitational potential of a classical massive source and
they do not trigger a decoherence of the neutrino state -- only the
hard real graviton emissions are relevant.

% \vspace{-0.4cm}
\subsection{Probability for quantum gravitational decoherence}
% \vspace{-0.2cm}
The cross section of the considering GBH process can be 
enhanced for the Galactic Center
($\sim 10^6 - 10^9$ Solar masses) or the dark matter halo
($\sim 10^{20} - 10^{24}$ Solar masses). It can also be enhanced for
ultra-relativistic neutrinos which are potentially detectable at
neutrino observatories such as IceCube and Super-K. As is our main
result, we notice that the GBH scattering may cause the quantum
decoherence of astrophysical neutrinos and this effect can be
measured via neutrino flavor composition measurements. A massive
classical source of the gravitational field may not necessarily be a
black hole, but any compact star or, in general, any bound
gravitational potential induced by continuous matter distribution in
the Galactic disk and Halo.

Due to rather large cross sections it can be that most of the
astrophysical neutrinos which are observed at the Earth from a given
direction and have passed in close vicinity of a massive object
would have experienced the quantum decoherence due to a
graviton-induced scattering. In other words, the probability for a
given neutrino in a superposition of mass states $f$ to decohere,
being ``transformed"
into one of the mass states $a=1,2$ or $3$ in the GBH process,
$P_{G}\sim |A^{(G)}|^2$, may be large for possible
(massive) astrophysical sources of classical gravitational fields,
depending on details of astrophysics and quasi-classical gravity. 
Deviations from quasi-classical gravity which illuminate the fundamental
quantum gravity theory may also be relevant. To parameterise
this we can define
\begin{equation} \label{P-G}
   P_G\equiv\frac{N_{\rm{G\nu}}}{N_{\rm{init}}} 
   \end{equation}
where $N_{\rm{G\nu}}$ is the number of neutrinos which have radiated
off an energetic graviton with $E_G\gtrsim \Delta m_{ij}$ while
being scattered off a massive object\footnote{On the way to the
Earth, a produced mass state may experience more graviton-induced
rescatterings in classical gravitational potentials which do not
affect the coherence of the neutrino state any longer, but may cause
an additional energy loss of the propagating neutrino into the
gravitational radiation.}, and $N_{\rm{init}}$ is the total number
of neutrinos which have been emitted off an astrophysical source. As
we will demonstrate below, the $P_G$ value can be measured via
neutrino flavor composition observations leading to a promising
opportunity for experimental tests of quantum gravity induced
interactions.

A precise theoretical calculation for $P_G$ is influenced by many
potentially relevant aspects. First, it depends on a quantum gravity
model through model-dependent local neutrino-graviton couplings thus
offering a good opportunity for experimental tests of quantum
gravity. Due to the extended nature of many of the sources, there might
be strong differences for models with some non-locality. 
Second, it may be influenced by yet unknown higher-order
corrections and by multiple rescatterings of a neutrino off a
massive source, multiple massive sources, or a diffuse source such
as the dark matter halo which the neutrino passes through on its
path to the Earth (in this case, the eikonal approximation for
neutrino-graviton rescattering can be used \cite{Kabat:1992tb}).
Thus, the actual cross sections may significantly vary depending on
environment a neutrino propagates in. Thirdly, the astrophysical
neutrino flavor composition may depend on production processes which
may currently be unknown. Also, energy loss of the
neutrino due to the hard Gravi-sstrahlung in each scattering
event should be taken into consideration, together with other effects which
change the coherence of the neutrino state. This could also be used to identify
a graviton interaction, if for example this lower energy flux comes some short time later
than the initial flux. Finally, including
possible dense astrophysical media might be important as the
neutrino may have additional weak rescatterings off normal matter
acting on the neutrino leaving the neutrino in a superposition
of mass eigenstates when it arrives at the Earth. Therefore, additional
astrophysical information is desired to constrain these
uncertainties. All of the above aspects are the major unknowns in
making predictions for the $P_G$ quantity which require a
further effort of the quantum gravity, neutrino and astrophysics
communities.\\

% \vspace{-0.4cm}
\subsection{Gravitational Compton scattering}
% \vspace{-0.2cm}

Another possibility for quasi-classical gravity induced interactions with
neutrino participation is shown in Figure~\ref{fig:BH-nu}(b). This is
the (tree level) gravitational Compton scattering of a neutrino off
a real graviton in cosmological medium.
The latter process has been previously studied in
Ref.~\cite{Voronov} and in many other papers. The cross section in
this case is always extremely small $\sigma\sim E_{\nu}^2/M_{Pl}^2$
for a MeV neutrino, and real graviton fluxes are not expected to
compensate for such a huge suppression. This process seems less
interesting when applied to astrophysical neutrino flavor
composition. Hypothetically, this effect could be considered in
exotic cases of ultra-relativistic neutrinos and/or in the very
early Universe where the graviton fluxes might have been rather
intense.
% \vspace{-0.4cm}

\section{Quasi-classical Gravity measurement proposal}
% \vspace{-0.2cm}
As presented above, the neutrino in a mass eigenstate does not
oscillate unless it scatters off ordinary matter via a weak channel
which will cause it to be in a flavor eigenstate. It is
likely that the $Z,W$-mediated scattering happens only in the
Earth-based detector enabling us to access information about the
graviton-neutrino scattering which might have happened far away from
the solar system. In the considered situation, the neutrino plays an
analogical role of an electric charge in a quantum measurement of
the microscopic Coulomb field properties. From the quantum
mechanical point of view, a massive object 
(e.g. dark matter distribution or the galactic center)
vicinity can then be viewed as a macroscopic ``detector'' of
gravitons. The neutrino scattering off a massive object and
radiating an energetic graviton by means of the local
graviton-neutrino coupling would be an elementary act
of quantum mechanical measurement, and the neutrino conveys the
quantum information about the act of graviton measurement to the
Earth (see figure \ref{fig:fig2diagram}). The neutrino does not undergo oscillation or demonstrate
properties consistent with being a superposition of mass eigenstates
since it is in a definitive mass state during the propagation and
graviton interaction. The neutrino not interacting weakly as
it travels is a good approximation due to extremely
weak interactions of neutrinos with ordinary matter. Then an
Earth-based detector will ``read off'' the results of the ``graviton
measurement'' which has taken place at the massive
object.

Previously, in Ref.~\cite{Rothman:2006fp}, it has been claimed that
it is not possible to detect a single graviton with a planet-scale
detector. Our proposal is to measure the described graviton-neutrino
scattering effect (specifically, the gravitational Bethe-Heitler
scattering of neutrino off a massive
object) experimentally, which is the best possibility for indirect
graviton detection proposed. Remarkably, we consider a super 
massive-scale ``detector'' of energetic gravitons, with
neutrinos serving as the most efficient carrier of the information
about such a measurement to the Earth.

\subsection{Quantum gravitational decoherence effect on neutrino
oscillations}

Here we consider a very massive source of strong gravitational
fields like a cluster of stars (for example, the center
of our Galaxy) or a dark matter halo as a good example of a graviton ``detector''. 
This Section provides predictions
for such an extreme large-scale quasi-classical gravity measurement.

As we have demonstrated above, the probability of an individual
(elementary) act of the ``quantum gravity measurement'' defined by
the graviton-neutrino cross section can be rather large due to a
large GBH cross section and there may be scenarios where it 
should not be neglected. Especially,
utilizing the dense region of stars and black holes in the Galactic Center (GC) as our ``graviton
detector'' in the above sense, one could expect that a significant
fraction of neutrinos passing by the dense region would have
experienced the GBH scattering. Then since many of the neutrinos are
now in a mass eigenstate, they will no longer undergo flavor
oscillation. Due to the neutrino existing in a mass eigenstate during 
propagation, further graviton re-scattering would not constitute
additional quantum measurements of an ``undetermined'' quantum state.
Depending on the astrophysical process, one might
favor energies of neutrinos where the neutrino
oscillation may not be suppressed due to the MSW effect where the neutrino
exists in a single mass eigenstate, so that the
graviton-induced effect would be cleaner. We suggest that
this effect could be tested in neutrino telescopes and observatories
by looking at the Galactic Center neutrino flavor composition and
comparing it to the composition expected without quantum gravitational
decoherence. It might be possible that close, ``standard candle'', neutrino
emitters in other parts of the sky provide a flux of neutrinos
which have not undergone quantum gravitational decoherence\footnote{Note, a
very similar effect should take place in flavor oscillations in the
neutral kaons $K_l,\,K_s$ system as well.}.

The general formula for the number of electron type
neutrinos observed from an electron type source in the vacuum is:
\begin{align}
    \frac{N_{\rm e,det}}{N_{\rm e,init}} \propto
     P_{\rm ee,\infty}^{\rm vac}(1-P_G) +
    P_{G}\sum_{i=1,2,3}V_{e i}V^*_{i e} V_{e i}V^*_{i e} \,.
\end{align}
Here $P_{{\rm ee},\infty}^{\rm vac}$ is the standard vacuum
oscillation probability \cite{Beringer:2012pdg} far away from the neutrino
source and $P_G$ is the probability for neutrino in a flavor state to interact
with at least one graviton (\ref{P-G}) which will depend on the
graviton-neutrino scattering cross section. Every mass
eigenstate of the (relativistic) neutrino shares the same energy so
$P_G$ takes the same value.

\begin{figure}[h!]
\centerline{\epsfig{file=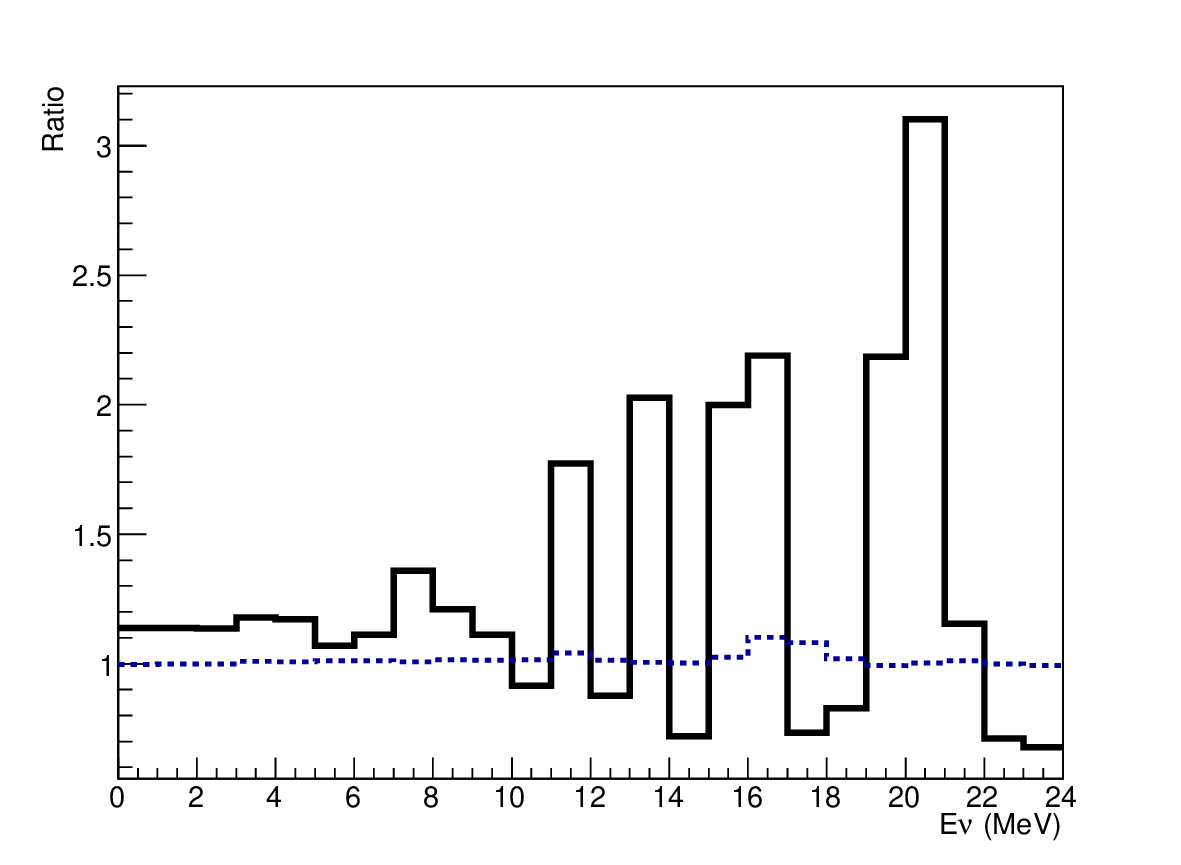,width=8cm,angle=0,scale=1.4}}
\caption{Here is plotted the difference due to the Earth matter effect
in the observed number of electrons (assuming an initial flux that
only contains electron neutrinos). The difference is defined to to be the ratio
of the observed number of electron type neutrinos for the case where there is a
maximal earth matter effect (impact parameter is $\sin\left(.576\right)$) and 
the case where there is no earth matter effect. Shown are two curves, black solid
for $P_G=0$ (propagation decoherence case) and blue dotted for the
$P_G=1$ (quantum decoherence case). This calculation was done with a
uniform distribution of decoherent electron neutrinos. The difference due to the
Earth matter effect may be higher than 300$\%$. In the case of fixed $P_G=1$, the flavor
composition of the neutrino flux is minimally affected by the Earth
at low energies.} 
\label{fig:nue}
\end{figure}

If all neutrinos have interacted with at least one graviton, i.e.
fixing $P_G=1$, than the expression for the total
$\nu_{e}\rightarrow \nu_{e}$ transition probability becomes
\begin{equation}
    P_{\rm ee}^G = \cos^4\theta_{12} \cos^4\theta_{13} +
    \cos^4\theta_{13} \sin^4\theta_{12} + \sin^4\theta_{13}
\end{equation}
where $\theta_{12}$, $\theta_{13}$, and $\theta_{23}$ are the
standard neutrino vacuum mixing angles. 
This basic formula is our prediction (in vacuum) for the ``maximal
decoherence'' scenario valid for $P_G\simeq 1$. In the standard
Large Mixing Angle (LMA) global fit with
$\sin^{2}\theta_{13}=0.025$, $\sin^{2}\theta_{12}=0.31$, and
$\sin^{2}\theta_{23}=0.60$ (but with $\delta_{CP}=0$)
\cite{GonzalezGarcia:2012sz}, the value for the transition
probability throughout a range of neutrino energies is shown in
Figure~\ref{fig:nue}. The difference between the predictions for
$P_{G}=1$ and $P_{G}=0$ is that for the $P_{G}=0$ is due to the neutrino
which has experienced propagation decoherence will have a constant phase
difference in the flavor basis at the earth/vacuum boundary giving
possibly an over than 100\% change in the survival probability depending on
neutrino energy, the detector resolution, and the detector location.
Further detail is available in appendix \ref{sec:appendixnote}. Here the simulation for
neutrino propagation in matter and vacuum was based on \cite{wimp}.

As one notices in Figure~\ref{fig:nue}, the relative effect ranges from
a few percent at $E_\nu\simeq 2$ MeV to about
$350$ \% at $E_\nu\simeq 20$ MeV. This is expected because the
electron type component of the super position of three decoherent
mass states is larger than the electron type component of the single
mass states. The Earth matter effect depends on neutrino energy and so
the effect will be larger at higher energy (20 MeV and above) than at smaller 
energies (2 MeV). This gives a quite noticeable difference between the case
where the neutrino undergoes {\it propagational decoherence} relative to the
case where the neutrino experienced {\it quantum decoherence}.

For illustration in this calculation we use a constant, maximal probability 
for quantum decoherence case, $P_G=1$. In practice we expect this to be 
less than one and to depend on the neutrino energy $E_\nu$. 
The $P_G(E_{\nu})$ value can be considered as an
observable and extracted from the flavor composition data and
further compared to theoretical calculations. Possible sources of
neutrinos in extreme astrophysical environments include the
aforementioned dense Galactic Center, but also ordinary stars, SNe, GRB, AGN, and
other galactic or extragalactic sources \cite{Strumia:2006db}\cite{Kim}. Quantum gravity models
should aim at predicting the $P_G$ in these extreme environments as
a function of astrophysical parameters and neutrino energy so that
favored models can be constrained by the neutrino flavor data.

The neutrino flux spectrum from astrophysical sources is still being
modeled \cite{Lai:2010tj,Esmaili:2009dz}. By
comparing observed neutrino flavor composition for neutrinos passing
through the Earth to that of neutrinos which have not passed through
the Earth the flux can be divided out and a possibly large effect may
be observed (for the above LMA global fit). This could be visible in
flavor data at large statistics.

This is our prediction for the quantum gravity-induced effect
on the detected flavor composition in the ``maximal quantum
decoherence'' scenario. While numerically the effect can be
very large, one would certainly need to have a good
understanding of all the other statistical and systematical
uncertainties for a possible measurement of the dependence of 
$P_G$ on neutrino energy. Current generation neutrino observatories
can observe approximately ten thousand total events from nearby
($\sim 10$ kpc) SNe, however, next generation neutrino observatories
(such as Hyper-K) can provide statistics to reduce
systematical uncertainties \cite{Scholberg:2012id}. Additionally,
improvements are needed in current neutrino flavor reconstruction
technologies, which can reconstruct the neutrino flavor to the
percentage level \cite{Akiri:2011dv}. It has been pointed out that
sensitive neutrino detectors should be placed in North America
(Sudbury), Japan (Hyper-K), Chile (ANDES), and Antarctica (Beyond
DeepCore) to best observe SNe neutrinos and the effect that the
Earth has on SNe neutrinos \cite{Machado:2012ee}.

\section{Conclusion}

In conclusion, we have considered a quasi-classical gravity process,
the gravitational Bethe-Heitler scattering of a neutrino off a
massive object accompanied by an energetic graviton radiation, which
can have a rather large cross section proportional to the mass
squared of the classical source. Due to hard gravitons interacting
with a neutrino mass eigenstate only, opposite to weak bosons which
interact with a flavor eigenstate only, the considered process is a
measurement of the incoming neutrino flavor state at the quantum
level, causing its decoherence in a different manner than the process
which can be called propagational decoherence or other sources of decoherence. This
quantum decoherence affects astrophysical neutrino behavior. Namely,
quantum decoherence can be considered as a specific
quasi-classical gravitational measurement of the neutrino propagating state,
which changes the behavior of the neutrino in the presence of a
potential (such as the Earth matter effect) compared to the traditional
source of decoherence known as propagation decoherence, which can be
observed in the neutrino flavor composition in an Earth based
detector (see appendix \ref{sec:appendixnote}).

This enables the utilization of neutrinos traveling across the
Galaxy as a source of information about the graviton-induced
interactions they might have experienced on their journey to Earth.
Specifically, the measured probability to find a given flavor
component in the neutrino flux coming from a vicinity of a super massive black
hole or another super massive object (galactic center 
or dark matter halo) will be different from the
corresponding probability measured from a source of neutrinos where
the neutrinos never pass near a massive system. In the case where
no astrophysical neutrinos can be identified which have not interacted
with a gravitational potential, the flavor composition can be compared
to the expectation for the Earth matter effect which can be determined
using reactor, atmospheric, and accelerator neutrinos. We have explicitly
demonstrated that the maximal difference corresponding to an
assumption that all of the detected neutrinos have experienced an
interaction with a graviton, i.e. $P_G=1$, is large and can be 
measurable at high statistics. Assuming that the required astrophysical 
conditions are met for large $P_G$, this would provide
a first measurement of quasi-classical gravity. Further discrimination of
quantum gravity models would require more statistics and detailed
calculations using these models. Additionally, we note that the energetic
graviton bremsstrahlung would cause a significant decrease in energy
of the neutrinos which are scattered at large angles. Since the galactic
center is not only massive but a source of neutrinos, it might be possible that
a large enough SNe in the galactic center would produce enough neutrinos
so that the existence of graviton bremsstrahlung could be induced by a group
of neutrinos arriving a short time after the initial group with a lower median energy.
This could be used in addition to the flavor to investigate quantum 
gravity models.

Thus, the probability for a neutrino state to interact with at least
one energetic graviton, $P_G$, is considered to be a new observable
containing information about the quantum gravity scattering process.
An estimate of the $P_G$ value from neutrino flavor composition data
with good angular resolution would provide an important experimental
test for quantum gravity models. This is the major proposal we make
in our paper. We do not expect $P_G\sim1$ in most scenarios
and a realistic theoretical estimate for $P_G$ depends on
many factors and is not well-constrained yet. For a distant enough 
source, there are many potential scatterers which may provide 
the maximal case of $P_G=1$.

The difference between propagation decoherence and quantum 
gravitational decoherence is a crucial component our study and so 
we provide short summary. In the classical case \cite{Beuthe:2001rc} 
the neutrino is produced and detected
in distinct flavor states (at the astrophysical source and the
earth detector) and exists as a quantum mechanical superposition
of mass states due the mass states being indistinguishable to the
detection process. If the neutrino passes near a massive object,
then it might undergo what we described as {\it classical decoherence}
\cite{Christian:2005qa}\cite{Diosi}\cite{Penrose:1996cv}. The different 
mass states continue to exist and
make up the neutrino even if they cease to overlap due to
what we describe as {\it propagation decoherence}\cite{Beuthe:2001rc}.

In the quasi-classical gravity case (this study), the neutrino is produced
and detected in distinct flavor (at the astrophysical source and the
earth detector), however, the neutrino exists in a single mass state
due to being ``observed'' by the emitted graviton which distinguishes
which mass state the neutrino exists in. We describe this effect as
{\it quantum decoherence}. Since only a single mass state exists, the
demonstrated phenomena are different such as that which is described
by the Earth matter effect where the electron type component of the
neutrino experiences the electromagnetic potential of the Earth
differently than the other components. For low energy neutrinos in a
single mass state, the Earth matter effect is less than 1\%; in
contrast to the Earth matter effect for a decoherent (due to
propagation) superposition of mass states which have a stronger
electron type neutrino component and experience a stronger Earth matter
effect depending on neutrino energy.

While we give an explicit calculation of the GBH process to
demonstrate that the emission of a hard graviton via gravitational
Bremsstrahlung is relatively large, and used this fact to
motivate discussion of a maximal possible signature, i.e. $P_G=1$,
we expect the calculation of gravitational Bremsstrahlung to require
corrections similar to that of photon Bremsstrahlung
\cite{Tseng,Amundsen} for an extended source in addition to the
loop-induced corrections for a full theory of quantum gravity. Additionally,
in many considered astrophysical scenarios, the astrophysical distances 
involved would cause $P_G$ to be small. This requires further work.

Having all that in mind, as a natural starting point in this very
first paper we would like to present the basic concept/idea of
quantum decoherence due to large angle neutrino-graviton
interactions (gravi-strahlung) in strong gravitational fields and
its possible effect on neutrino flavor observables. In this paper we
report on our preliminary study of such a graviton-induced effect on
neutrino oscillations and motivate future studies in this direction.
We plan to improve our simulation with fluxes and the astrophysical
medium in a future study. The possibility that $P_G$ is not zero in
the vicinity of the Sun should be considered as well. For example,
using the same simulation as used to produce Figure~\ref{fig:nue} we
find a preliminary effect for Solar neutrino of approximately $3\%$ for
$P_G=0$ in the integrated B$_8$ spectrum while we see an asymmetry of
$\sim0\%$ for $P_G=1$. Explicitly, the length and energy dependence
of neutrino flavor oscillation will depend on the relative strengths
of the graviton-neutrino interactions, the matter properties, and
the vacuum oscillation parameters. Inclusion of these possibilities
in the global neutrino oscillation parameter fit will be left for a
later paper. Additionally, extragalactic neutrinos should be
considered with additional care as GBH scattering of the neutrino
off the diffuse dark matter Halo may play a role. Finally, the issue
of coherent production of neutrinos is not considered in this study
and should be studied in detail in a future work.

The program used to produce Figure~\ref{fig:nue} can be found on the arxiv
\cite{thiswork}. For further information about semi-classical quantum gravity
see \cite{Paszko:2010zz} and for further information about graviton bremsstrahlung see
\cite{Gould,Barker:1974hf,Peters:1970mx}.

The authors declare that there is no conflict of interests regarding the publication of this paper.

In the review process, \cite{Ahluwalia:1998xb} was brought to the attention of the authors. Here the idea that a neutrino in a superposition of mass eigenstates may be projected to a single mass eigenstate by a gravitational couplings was presented.

\acknowledgments
%\section{Acknowledgments}

%%\vspace{0.1cm} {\bf Acknowledgments}.
Stimulating discussions and
helpful correspondence with Sabine Hossenfelder and Alexei Vladimirov
are gratefully acknowledged. 
%This work was supported in part
%by the Crafoord Foundation (Grant No. 20120520). 
J. M. was supported
in part by PROYECTO BASAL FB 0821 CCTVal and Fondecyt (Grant No. 11130133). 
R. P. was supported in part by Fondecyt (Grant No. 1090291) and the 
Crafoord Foundation (Grant No. 20120520). R. P. is grateful to the
``Beyond the LHC'' Program at Nordita (Stockholm) for support and
hospitality during a portion of this work. This research was supported in part by the 
National Science Foundation under Grant No. NSF PHY11-25915. 

\appendix

\section{Note on measuring graviton induced decoherence}
\label{sec:appendixnote}

The central effect of the graviton observation of the neutrino which is utilised in this proposed measurement
is that a superposition is different than a classical ensemble of states. Distinguishing these two things is 
of key interest to the Quantum Information and Quantum Foundations communities and they have been shown
to be different in experiments which investigate Bell Inequalities. To quote a member of the Quantum Foundations
community who are also interested in distinguishing the situation where the particle is in a superposition (neutrino
which has not undergone an interaction with a graviton, in our case) and those where it has and the wave function has collapsed
(in the mass basis in our case, where the neutrino has undergone an interaction with a graviton) \cite{Schlosshauer:2003zy}:

"It is a well-known and important property of quantum mechanics that a superposition of states is fundamentally different from a classical ensemble of states, where the system actually is in only one of the states but we simply do not know in which (this is often referred to as an {\it ignorance-interpretable,} or {\it proper} ensemble)."

What is required to distinguish these two cases is for the phase between the states to not be rapidly varying. In the case where 
interference phenomena may be observed (the phenomena of neutrino oscillation for neutrinos) this is obviously the case. For the case 
where the particle is still coherent but the phase difference between states is rapidly varying, it is obvious that it is impossible 
to differentiate a classical ensemble from a superposition. For neutrinos this is the situation where there is still overlap between 
the states but the energy resolution of the detector is not good enough to observe the oscillation, and has been talked about in
\cite{Farzan:2008eg}\cite{Beuthe:2001rc}.

However, there is an additional case where the neutrino in the flavor basis ceases to oscillate. The states no longer overlap. This 
is the case of propagation decoherence and the generally the case for astrophysical neutrinos. In this case, in the flavor basis, the 
neutrino has a constant phase difference between (matter) states. If we measure this state without making any changes to it based
on the phase difference we get the same result as if we measure a classical ensemble of states (the quantum gravity decoherence 
case). However, if we modify this (flavor) state by sending it through matter, the constant phase difference is changed differently than 
the classical ensemble of states and the neutrino can be distinguished as being in a classical ensemble of states (or having undergone 
quantum gravity decoherence) rather than a separated superposition. The boundary between matter regimes has a finite width and so
the (flavor) state is going to have a constant phase difference (for large enough separations), independent of the energy resolution of
the final (flavor) detector which collapses the wave function.

This can be clearly described in the density matrix formalism. In the formalism, the evolution of the density matrix is given by
\begin{equation}
\dot{\rho}=-i[ H,\rho]
\end{equation}
where $\rho$ is the density matrix and $H$ is the hamiltonian. It has been shown \cite{Stodolsky:1998tc} that if $\dot{\rho}=0$, knowledge about the particulars
of the wave packet is unnecessary and as a consequence you can not distinguish wave packet separation from the case where you have a measurement 
at large distances (or you have a graviton interaction leaving the neutrino in a distinct mass eigenstate). 
However, we are considering the case where wave packet separation has occurred and then the neutrino passes through jump in the potential. 
We can describe change of basis from flavor to vacuum as $C_1$
and the change of basis from flavor to matter as $C_2$. We can describe the decoherence process as $D$, which nullifies the off diagonal components
of the density matrix (for example \cite{Akhmedov:2014ssa}). Note that $D$ does not commute with $C$. We can also describe the adiabatic process of the neutrino passing through the earth as
$H_m$. We then consider 
\begin{equation}
[C_2^{-1} H_m C_2 C_1^{-1} D C_1, \rho]
\end{equation}
and note that it is not $0$. Thus $\dot{\rho}\neq0$ and wave packet information is relevant and furthermore, as presented in this study, you may distinguish 
wave packet separation from graviton induced decoherence.

For the case of 2 neutrinos flavours, the constant phase difference for a neutrino in a separated superposition is obvious and is given by $\frac{\pi}{2}$.
For the case of 3 neutrino flavours, it is more difficult, and by brute force phase differences which work for the neutrino parameters used in this study are $0.756253i$ and $1.477224i$, the results of which are plotted in Fig \ref{fig:nue}.

%===========================

%===========================

\end{document}